\documentclass[multphys,vecphys]{svmult}

\usepackage{makeidx}
\usepackage{graphicx}

\usepackage{multicol}
\usepackage[bottom]{footmisc}

\makeindex

\begin{document}

\title*{A simple model for quasar density evolution}

\author{Hannes Horst\inst{1,2}\and
Wolfgang J. Duschl\inst{2,3}}

\institute{European Southern Observatory, Casilla 19001, Santiago
19, Chile \texttt{hhorst@eso.org} \and Institut f\"ur Theoretische
Astrophysik, Albert-Ueberle-Str. 2, 69120 Heidelberg, Germany
\texttt{wjd@ita.uni-heidelberg.de} \and Steward Observatory, The
University of Arizona, 933 N. Cherry Ave, Tucson, AZ 85721, USA}

\maketitle

\begin{abstract}
It is widely agreed upon that AGN and Quasars are driven by gas
accretion onto a supermassive black hole. The origin of the latter
however still remains an open question. In this work we present the
results of an extremely simple cosmological model combined with an
evolutionary scenario in which both the formation of the black hole
as well as the gas accretion onto it are triggered by major mergers
of gas-rich galaxies. Despite its very generous approximations our
model reproduces the quasar density evolution in remarkable
agreement with observations.
\end{abstract}

\section{Introduction}
\label{horst02intro}

While it is widely agreed upon that galaxy-galaxy interactions and,
in particular, mergers play a crucial role for the growth of
supermassive black holes in quasars and also for providing the fuel
for quasar activity, there is still dispute whether these black
holes are of primordial nature or not. N-body simulations (e.g. by
\cite{barnes96}) show that the tidal forces in interacting galaxies
can trigger strong gas inflows towards the center of the merger. The
mass of this gas is sufficient to build a supermassive BH of $10^{7}
\dots 10^{10} \textrm{M}_{\odot}$ and provide enough fuel for quasar
activity. On the basis of the $\beta$-viscosity model by
\cite{duschl00} calculations by \cite{duschl06} show that it takes
less than $10^{9}\,$a from the merger to form a fully developed
quasar even if no primordial supermassive BH was originally present.
This model can be tested by comparing the resulting co-moving space
density of quasars to those derived from observations (e.g.
\cite{silverman05}).

\section{Outline of the model}
\label{horst02bhs}

Based on results by \cite{duschl06} we assume an average time delay
of $5 \cdot 10^8$a between the merger and the peak quasar activity.
The individual values of this delay depend, of course, on details
like the size ($s_{\textrm{{\scriptsize disk}}}$) and initial mass
($M_{\textrm{{\scriptsize disk}}}$) of the disk. Estimates of the
relevant (viscous) timescale show its dependence $\propto
s_{\textrm{{\scriptsize disk}}}^{3/2} M_{\textrm{{\scriptsize disk,
i}}}^{1/2}$. This leads to the -- at first glance
surprising---finding of the faster formation of the more massive
black holes. For our present purpose, however, an average value of
this delay serves its purpose. For more details of this model, we
refer the reader to the contribution by Duschl and Strittmatter in
this volume and in \cite{duschl06}.

For our purpose it is sufficient to use a ``test universe" comprised
of 50 000 galaxies. This test universe is expanding according to an
Einstein-de Sitter-cosmology with a Hubble constant of $H_{0} = 72
\frac{\textrm{{\scriptsize km}}}{\textrm{{\scriptsize s}} \cdot
\textrm{{\scriptsize Mpc}}}$. Our galaxies are treated as particles
with a finite cross section and a thermal velocity dispersion of
$v_{s} = 300 \frac{\textrm{{\scriptsize km}}}{{\scriptsize
\textrm{s}}}$. In this framework we compute the merger rate for each
simulated time bin ($10 \cdot 10^6\,$a) by taking into account two
different processes: Direct geometrical hits and gravitationally
driven mergers. For the first process we assume every galaxy to have
a spherical cross section with a radius of 15 kpc. For the second
process the cross section radius is $r_{\textrm{{\scriptsize cs}}} =
2 G^{*} M v_{\textrm{{\scriptsize rel}}}^{-2}$, with $M$ being the
mass of one galaxy and $v_{\textrm{{\scriptsize rel}}}$ being the
relative velocity between both. We assume a delay between the merger
and the onset of the quasar phase of $5 \cdot 10^8\,$a for the
geometrical hit and $1 \cdot 10^9\,$a for the gravitationally driven
merger. The quasar phase in turn is assumed to last for another $5
\cdot 10^8\,$a.

\begin{figure}
  \begin{center}
    \includegraphics[width=\textwidth]{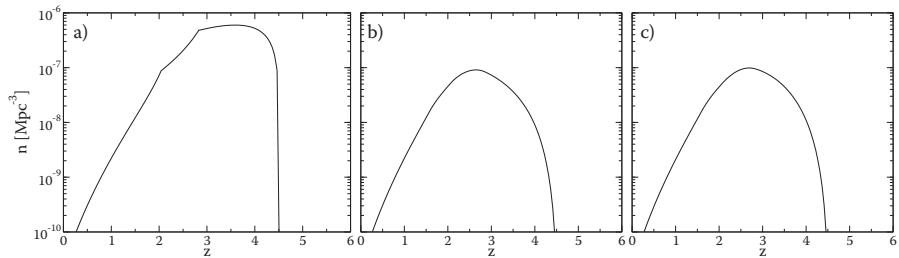}
  \end{center}
  \caption{Co-moving space density of quasars as computed by our
             simulation. In panel a) all galaxies are in place at
             the start of the simulation. In panels b) and c) the
             number of galaxies is gradually increased over the first
             $5 \cdot 10^8$a. In panel b) a constant galaxy formation
             rate and in panel c) a Gaussian formation rate were used
             -- the resulting co-moving space density of quasars is
             almost exactly the same.}
  \label{horst02simul}
\end{figure}

In our model the Universe is treated in very simple manner: As an
expanding spherical box. The size of this box is determined by
deriving today's matter density from the assumed Hubble constant
($\left(\frac{8}{3}\pi G^{*}\rho_{0} \right)^{1/2} = H_{0}$) and
then using the Friedmann-Lema\^{i}tre-equation to calculate the
according radius at any given time. We completely neglect structure
formation and start our simulation with large galaxies ($M = 10^{11}
M_{\odot}$) already in place $2 \cdot 10^8$a after the Big Bang. To
account for the formation of galaxies we increase their number over
the first $5 \cdot 10^8\,$a of our simulation until the final number
of 50 000 is reached (see fig. \ref{horst02simul} for the effect of
this procedure).

\section{Results}
\label{horst02results}

\begin{figure}
  \begin{center}
    \includegraphics[width=0.75\textwidth]{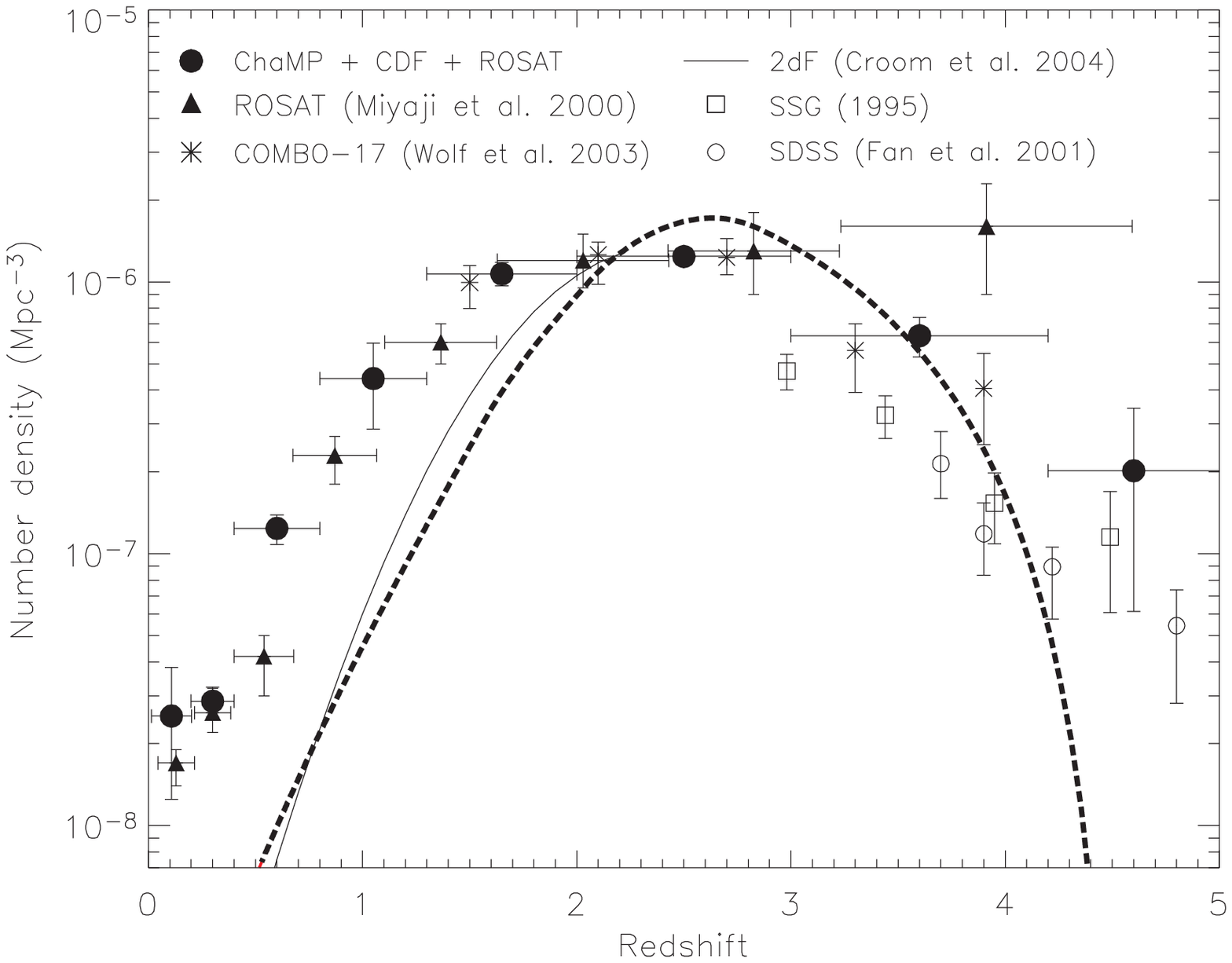}
  \end{center}
  \caption{Comparison of our simulated quasar density evolution with
             observational results. The broken line is the same as
             in panel b) of Fig. \ref{horst02simul}, while the
             underlying graph has been taken from \cite{silverman05}.
             The theoretical curve has been scaled to match the 2dF
             results.}
  \label{horst02obs}
\end{figure}

The comparison between our model results and observational data
compiled by \cite{silverman05} is shown in fig.\ref{horst02obs}. The
quasar density evolution at lower redshifts and the position of the
peak at $z \approx 2.5$ match very well. Please note that the X-ray
selected sample contains lower luminosity AGN in addition to
quasars. In this respect it is natural that our results resemble the
2dF curve (from \cite{croom04}) better than the data from
\cite{silverman05}. At higher redshifts the deviation between
simulated and observed co-moving spatial density is increasing. In
our model we clearly miss the earliest quasars which arise from
exceptionally fast evolving mergers. Despite this shortcoming of our
results their overall agreement with observational data is
remarkable.

\index{Accretion disk, AGN, cosmology, quasars, quasar evolution}

\printindex

\end{document}